\documentclass[useAMS,usenatbib,hyperref]{mn2e}
\usepackage[dvips]{graphicx}
\usepackage{subfigure}
\usepackage{color}
\usepackage{overpic}

\newcommand{\mnras}{MNRAS}
\newcommand{\aj}{AJ}
\newcommand{\aoms}{AoMS}
\newcommand{\apj}{ApJ}
\newcommand{\apjl}{ApJL}
\newcommand{\apjs}{ApJS}
\newcommand{\aap}{A\&A}
\newcommand{\araa}{ARA\&A}

\newcommand{\qjras}{QJRAS}
\newcommand{\pasp}{PASP}

\newcommand{\nar}{NAR}

\long\def\symbolfootnote[#1]#2{\begingroup%
\def\thefootnote{\fnsymbol{footnote}}\footnote[#1]{#2}\endgroup}
\renewcommand{\thefootnote}{\fnsymbol{footnote}}

\title[H-ATLAS: The FIR Properties of BAL Quasars]{{\em Herschel}*-ATLAS: the far-infrared properties and star-formation rates of broad absorption line quasi-stellar objects}
\author[J. M. Cao Orjales et al.]
{J. M. Cao Orjales$^1$\dag,
J. A. Stevens$^{1}$, M. J. Jarvis$^{1, 2}$, D. J. B. Smith$^{1}$, \newauthor M. J. Hardcastle$^{1}$,
  R. Auld$^{3}$,  M. Baes$^{4}$, A. Cava$^{5}$, D. L. Clements$^{6}$, \newauthor A. Cooray$^{7}$, K. Coppin$^{8}$,
  A. Dariush$^{6}$, G. De Zotti$^{9, 10}$, L. Dunne$^{11}$, \newauthor S. Dye$^{12}$, S. Eales$^{3}$, R. Hopwood$^{6, 13}$,
  C. Hoyos$^{12}$, E. Ibar$^{14}$, \newauthor R. J. Ivison$^{14, 15}$, S. Maddox$^{11}$, M. J. Page$^{16}$ and E. Valiante$^{3}$ 
\\$^{1}$Centre for Astrophysics Research, Science \& Technology Research Institute, University of Hertfordshire, Hatfield, AL10 9AB, UK
\\$^{2}$Physics Department, University of the Western Cape, Cape Town, 7535, South Africa
\\$^{3}$School of Physics and Astronomy, Cardiff University, The Parade, Cardiff, CF24 3AA, UK
\\$^{4}$Sterrenkundig Observatorium, Universiteit Gent, Krijgslaan 281--S9, Gent, 9000, Belgium
\\$^{5}$Departamento de Astrof\'{\i}sica, Facultad de CC. F\'{\i}sicas, Universidad Complutense de Madrid, E-28040 Madrid, Spain
\\$^{6}$Physics Department, Imperial College London, Prince Consort Road, London SW7 2AZ, UK
\\$^{7}$Department of Physics and Astronomy, University of California, Irvine, CA 92697, USA
\\$^{8}$Department of Physics, McGill University, 3600 rue University, Montr\'{e}al, Qu\'{e}bec, H3A 2T8, Canada
\\$^{9}$INAF -- Osservatorio Astronomico di Padova, Vicolo dell'Osservatorio 5, I-35122 Padova, Italy
\\$^{10}$SISSA, Via Bonomea 265, I-34136, Trieste, Italy
\\$^{11}$Dept of Physics and Astronomy, University of Canterbury, Private Bag 4800, Christchurch 8140, New Zealand
\\$^{12}$School of Physics and Astronomy, University of Nottingham, University Park, Nottingham NG7 2RD, UK
\\$^{13}$Department of Physical Sciences, The Open University, Milton Keynes MK7 6AA, UK
\\$^{14}$UK Astronomy Technology Centre, Science and Technology Facilities Council, Royal Observatory, Blackford Hill, Edinburgh EH9 3HJ
\\$^{15}$Institute for Astronomy, University of Edinburgh, Blackford Hill, Edinburgh EH9 3HJ, UK
\\$^{16}$University College London, Mullard Space Science Laboratory, Holmbury St. Mary, Dorking, Surrey, RH5 6NT, UK}

\date{Released 2012 Xxxxx XX}

\pagerange{\pageref{firstpage}--\pageref{lastpage}} \pubyear{2012}

\def\LaTeX{L\kern-.36em\raise.3ex\hbox{a}\kern-.15em
    T\kern-.1667em\lower.7ex\hbox{E}\kern-.125emX}

\begin{document}

\label{firstpage}

\maketitle
\begin{abstract}
We have used data from the \textit{Herschel}-ATLAS at 250, 350 and 500~$\mu$m to determine the far-infrared (FIR) properties of 50 Broad Absorption Line Quasars (BAL QSOs). Our sample contains 49 high-ionization BAL QSOs (HiBALs) and 1 low-ionization BAL QSO (LoBAL) which are compared against a sample of 329 non-BAL QSOs. These samples are matched over the redshift range $1.5\leq z < 2.3$ and in absolute $i$-band magnitude over the range $-28 \leq$ M$_{i} \leq -24$. Of these, 3 BAL QSOs (HiBALs) and 27 non-BAL QSOs are detected at the $>$5\,$\sigma$ level. We calculate star-formation rates (SFR) for our individually detected HiBAL QSOs and the non-detected LoBAL QSO as well as average SFRs for the BAL and non-BAL QSO samples based on stacking the \textit{Herschel} data.  We find no difference between the HiBAL and non-BAL QSO samples in the FIR, even when separated based on differing BAL QSO classifications. Using Mrk 231 as a template, the weighted mean SFR is estimated to be $\approx240\pm21$~M$_{\odot}$\,yr$^{-1}$ for the full sample, although this figure should be treated as an upper limit if AGN-heated dust makes a contribution to the FIR emission. Despite tentative claims in the literature, we do not find a dependence of {\sc C\,iv} equivalent width on FIR emission, suggesting that the strength of any outflow in these objects is not linked to their FIR output. These results strongly suggest that BAL QSOs (more specifically HiBALs) can be accommodated within a simple AGN unified scheme in which our line-of-sight to the nucleus intersects outflowing material. Models in which HiBALs are caught towards the end of a period of enhanced spheroid and black-hole growth, during which a wind terminates the star-formation activity, are not supported by the observed FIR properties.
\end{abstract}
\begin{keywords}
galaxies: active -- galaxies: infrared -- galaxies: photometry -- quasars:absorption lines 
\end{keywords}

\newpage
\section{Introduction}

Mergers and other interactions are thought to trigger AGN which can lead to outflows that affect the chemical make-up of the interstellar and intergalactic media, and limit star formation in the host galaxy (\citealt{Bower2006, Croton2006, DeLucia2007, Booth2009}) and possibly in the large-scale environment (\citealt{MJ2004, McCarthy2010}). The role of these outflows has become an important feature of contemporary models of the formation and evolution of galaxies over cosmic time, being required in theoretical attempts to understand galaxy `downsizing' (\citealt{Cowie1996, Scanna2005}). In semi-analytic models of galaxy assembly (\citealt{Granato2004}) AGN feedback removes dense gas from the centres of galaxies, heating up the surrounding Intergalactic Medium (IGM), while at the same time enriching the IGM with metals. Studies such as that of \citet{Gabel2006} indicate that outflows have high metallicities which could serve to enrich the IGM. Due to the IGM's low density, however, it cannot cool efficiently and so cannot fall back to the galaxy to fuel star formation.

Simulations of galaxy mergers (\citealt{Springel2005}) show that AGN feedback can heat the Interstellar Medium (ISM) and inhibits star formation, with outflows providing the energy and momentum feedback for the ISM of the host galaxy. Efforts to include these mechanisms (e.g., \citealt{Croton2006, Sijacki2007}) have focussed on two modes: a `radio mode' whereby a relativistic jet can heat interstellar gas and intracluster media, stopping further infall of mass (\citealt{Best2007, Best2012}), and a `quasar mode', which serves to stop star formation with an outflow of greater mass but with a lower velocity driven by the radiation from the quasar. At the same time, the quasar mode removes any leftover gas that might serve to enshroud the galaxy, turning it into a `classical' Quasi-Stellar Object (QSO). Recent work finds that when feedback is incorporated, the $\Lambda$CDM model provides a good fit to observed galaxy stellar-mass fractions and the $K$-band luminosity function (\citealt{Bower2006, Cirasuolo2010}). However, if feedback is excluded in galaxy group simulations, the temperature profiles are highly peaked and in disagreement with many of the observed properties of galaxies. The simulations also suffer from the well known overcooling problem, the resulting stellar mass fraction being several times larger than observed (\citealt{McCarthy2010}).

\symbolfootnote[0]{$^*${{\it Herschel} is an ESA space observatory with science instruments provided by European-led Principal Investigator consortia and with important participation from NASA. The H-ATLAS website is http://www/h-atlas.org/.}}
\symbolfootnote[0]{$^\dag${E-mail:jc10acd@herts.ac.uk}}

Gas outflows from AGN are primarily detected in X-ray and ultraviolet absorption against the inner portions of the accretion disk and/or the more extended broad-line region. Indicators for the origin of an absorption line system are (1) velocity width, (2) partial coverage, (3) time variability, and (4) high metallicity.
Historically, the criterion of velocity width has led to the use of three categories of intrinsic absorbers. Those with the greatest velocity dispersions are termed `Broad Absorption Line' Quasars, or BAL QSOs (e.g \citealt{Weymann1991}). These have absorption lines with FWHM greater than 2000~km\,s$^{-1}$. Those with very narrow absorption lines are known as `Narrow Absorption Line' QSOs, (NAL QSOs; e.g. \citealt{HamFer1999}). These have FWHM of less than 500~km\,s$^{-1}$. Finally the last type are known as Mini-BALs. These are those intrinsic absorbers whose velocity widths are between 500 and 2000~km\,s$^{-1}$ (e.g., \citealt{Hamann1997a}, \citealt{Churchill1999}).

There is still some debate in the literature regarding BAL QSOs, with the observed fraction ($\sim15$ per cent) being attributed to orientation, an evolutionary phase or some mixture of both.  In the orientation hypothesis, BAL outflows are present in all QSOs but they are only viewed as BAL QSOs when our line-of-sight intersects the solid angle subtended by the BAL outflow. In this situation, high column density accretion disk winds are accelerated by radiation pressure (\citealt{Murray1995, MurChi1998, Elvis2000}). This model attributes the fraction of QSOs to the fractional solid angle coverage of the BAL regions. Furthermore, it is able to account for the similarity of the continua and emission line features in BAL QSO and non-BAL QSO spectra once reddening is taken into account (\citealt{Weymann1991, Reichard2003}) and fits well with unified models (\citealt{Antonucci, Elvis2000}). Mid--infrared and far--infrared studies also show no difference in emission, consistent with there being similar dust masses within the host galaxy and BAL QSOs being an orientation effect (\citealt{Gallagher2007, Lazarova2012}). However, spectropolarimetry studies (\citealt{Ogle1999, DiPompeo2010, DiPompeo2011}) indicate that BAL QSOs are seen at a wide range of inclinations, which suggests that BAL QSOs may not be just a simple orientation effect.

The evolutionary scenario is rather different. In such a model, BAL QSOs are young QSOs that may have recently undergone a merger/starburst, and are observed as BAL QSOs when two special conditions are met; firstly, the QSO luminosity has reached a sufficient strength to accelerate gas to several thousand km\,s$^{-1}$ and secondly, there is a large mass of diffuse gas and dust in the nuclear regions leading to a rapid mass-loss phase. These BAL outflows could also play a role in terminating star formation and removing much of the obscuring dust and gas leading to a classical optically and soft X-ray luminous QSO (\citealt{Voit1993, Hall2002, Page2004, Page2011}). If BAL QSOs are caught towards the end of an epoch of enhanced star-formation activity then high dust masses within the host galaxies of BAL QSOs will yield higher average FIR flux densities. Since far-infrared emission is optically thin, there should be no dependence on the source orientation. This difference in the FIR properties in these two hypotheses can in principle be used to distinguish between them.

\citet{Omont1996} and \citet{Carilli2001}, both working at frequencies around 250~GHz, showed that there was weak evidence that BAL QSOs were more luminous at millimetre wavelengths than ordinary QSOs. This would favour an evolutionary scenario, but the work did not concentrate exclusively on the BAL QSO phenomenon, or compare star formation rates, and the samples were small (4 and 7 BAL QSOs respectively). Other work by \citet{Lewis2003}, \citet{Willott2003} and \citet{Priddey2007} was specifically aimed at improving our understanding of the BAL phenomenon, with larger samples of 7, 30 (41 non-BAL QSOs for a comparison sample) and 15 BAL QSOs (supplemented by data from \citealt{Priddey2003}) respectively. All of these samples were observed using the SCUBA camera on the James Clerk Maxwell Telescope at wavelengths of 450 and 850~$\mu$m. The samples of \cite{Willott2003} and \citet{Priddey2007} are composites of previous work by the authors at FIR wavelengths, and therefore have non-uniform sensitivity coverage. None of these studies find strong evidence for BAL QSOs to be more submillimetre luminous than non-BAL QSOs, a finding consistent with the orientation model for BAL activity. Within the BAL QSO population itself, however, \citet{Priddey2007} find tentative evidence for a dependence of submillimetre emission on C{\sc\,iv} absorption--line equivalent width and discuss the implications within the framework of simple evolutionary models.

In this paper we use {\em Herschel\/} (\citealt{Herschdata}) data to study the FIR properties of matched samples of BAL and non-BAL QSOs allowing us to build on previous work with improved statistics, better wavelength coverage around the FIR dust peak and better matching of the control sample. Throughout this work we use a cosmology where $H_0 = 70$~km\,s$^{-1}$\,Mpc$^{-1}$, $\Omega_{\rm m} = 0.3$, and $\Omega_{\Lambda} = 0.7$.

\section{The definition of a BAL QSO}

BAL QSOs have broad absorption lines which arise from resonance line absorption in gas outflowing with velocities up to $0.1c$ (\citealt{Weymann1991, Arav:2001lr, Hall2002, Reichard2003}). 
As a subclass of AGN, BALs until recently were thought to make up around $10-15$ per cent of the sources in QSO surveys (\citealt{Weymann1991, Tolea2002, Reichard2003}). However, \citet{Trump2006} quoted a much higher BAL fraction, i.e., an observed fraction within the 3rd Data Release of the SDSS (\citealt{Schneider2005}) of 26 per cent. It is possible that selection effects may bias us against BAL identifications, meaning that the intrinsic fraction is far higher than observed (\citealt{Hew2003, Dai2008}). 
Perhaps the most widely accepted classification scheme for whether a QSO is a BAL QSO is the Balnicity Index or $BI$, defined by \citet{Weymann1991},
\begin{equation}\label{eq:1}
BI=\int_{3000}^{25000}(1-\frac{f(-v)}{0.9})C\,dv .
\end{equation}
Here $f(v)$ is the continuum-normalized spectral flux at a velocity $v$ (in km\,s$^{-1}$) from the C{\sc\,iv} line rest wavelength 1549\,\AA\ (in the system frame). The dimensionless value $C$ is 0 unless the observed spectrum has fallen at least 10 per cent below the continuum for a velocity width of at least 2000~km\,s$^{-1}$ in the absorption trough, at which point $C$ is set to 1. Traditional BALs are defined to have $BI > 0$.

We construct our BAL sample using BAL QSOs selected using the $BI$ metric of equation \ref{eq:1} and a version of the metric which is `extended' as in \citet{Gib2009}, the lower limit of the $BI$ being 0~km\,s$^{-1}$.

BAL QSOs can be split into subclasses based on their observed spectral features. The first of these classifications are High-ionization Broad Absorption Line (HiBAL) QSOs. These objects contain absorption in Ly\,{\sc$\alpha\,\lambda1216$\,\AA}, N{\sc\,v}\,$\lambda1240$\,\AA, Si{\sc\,iv}\,$\lambda1394$\,\AA\ and C{\sc\,iv}\,$\lambda1549$\,\AA. They are the most prevalent sub-population, making up around 85 per cent of BALs. The Low-ionization Broad Absorption Line (LoBAL) QSOs contain all of the previously mentioned absorption features seen in HiBALs, but also contain absorption features in Mg{\sc\,ii}\,$\lambda2799$\,\AA\ and other low ionization species. These objects comprise around 15 per cent of BALs (\citealt{SprayFoltz1992} estimated 17 per cent). HiBALs and LoBALs are found to have redder continua than non-BALs (\citealt{Brotherton2001, Reich2003, Scaringi2009}), with LoBALs also being significantly redder than HiBALs (\citealt{Weymann1991, SprayFoltz1992}). The third and final class are FeLoBALs, which show the absorption features of LoBALs as well as absorption features arising from metastable excited levels of iron. The effects of these definitions on each measured variable are discussed in greater detail in Section 4. 

\section{Data and Sample Selection}

The BAL QSO sample that we use in this paper is a small proportion of those objects identified as a BAL QSO within the SDSS QSO Catalogue (\citealt{Gib2009}), which number 5039 BAL QSOs. By using the SDSS catalogue (both for BAL and non-BAL samples) a redshift of $z < 2.3$ provides an upper redshift limit where the SDSS colour selection criteria are relatively unaffected by BAL absorption (\citealt{Gib2009}). We use absolute magnitude in the SDSS $i$ band since it is less affected by dust reddening than the bluer bands.

As \citet{Gib2009} used the $BI$ as a classification mechanism, it is possible that some BALs will have been missed from the sample, and others included within the catalogue will be mis-classifications (though the strictness of the $BI$ should decrease the number). 

\begin{figure}
\centering
\subfigure{
\includegraphics[scale=0.45]{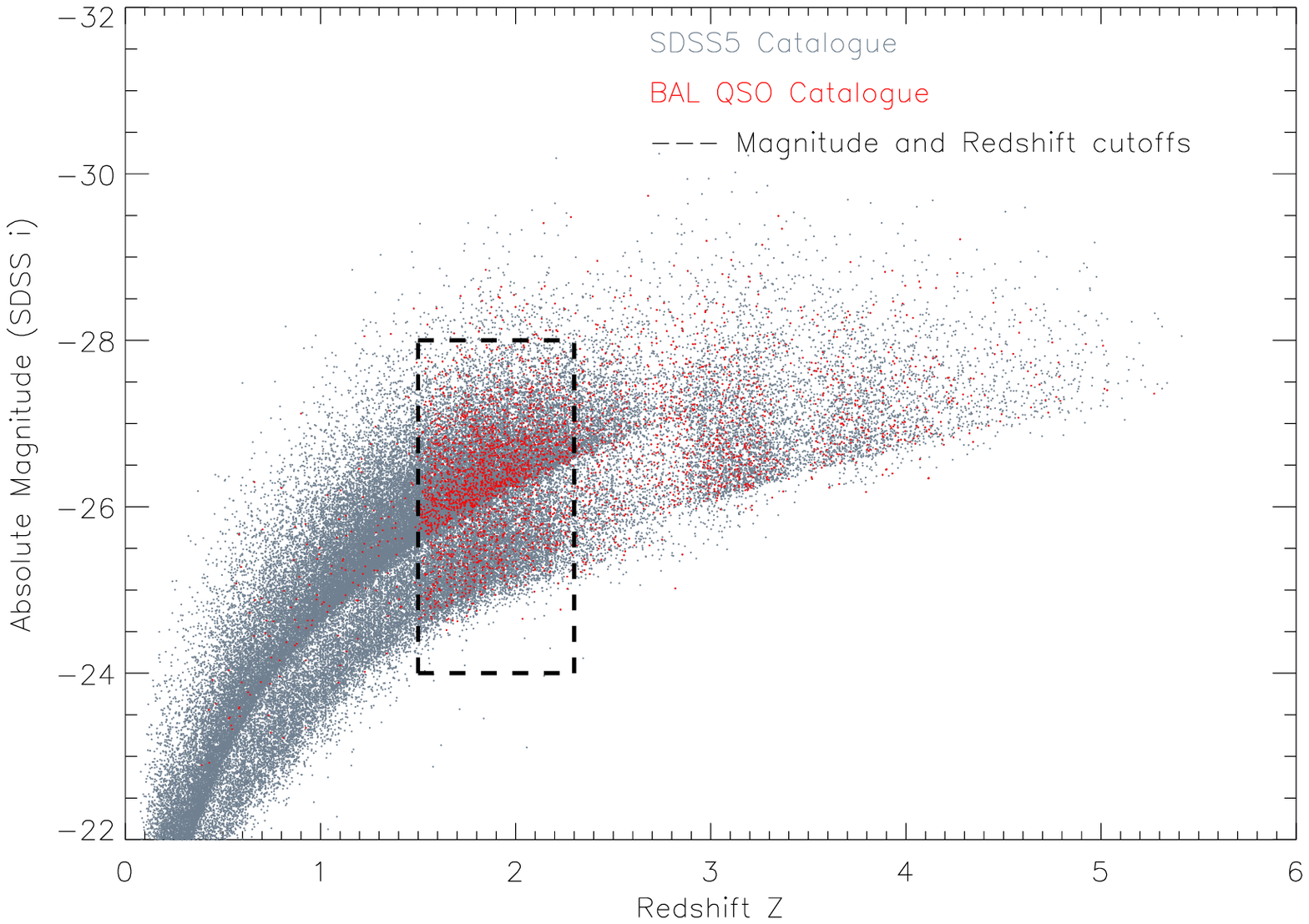} 
\label{fig:subabz}
}
\subfigure{
\includegraphics[scale=0.45]{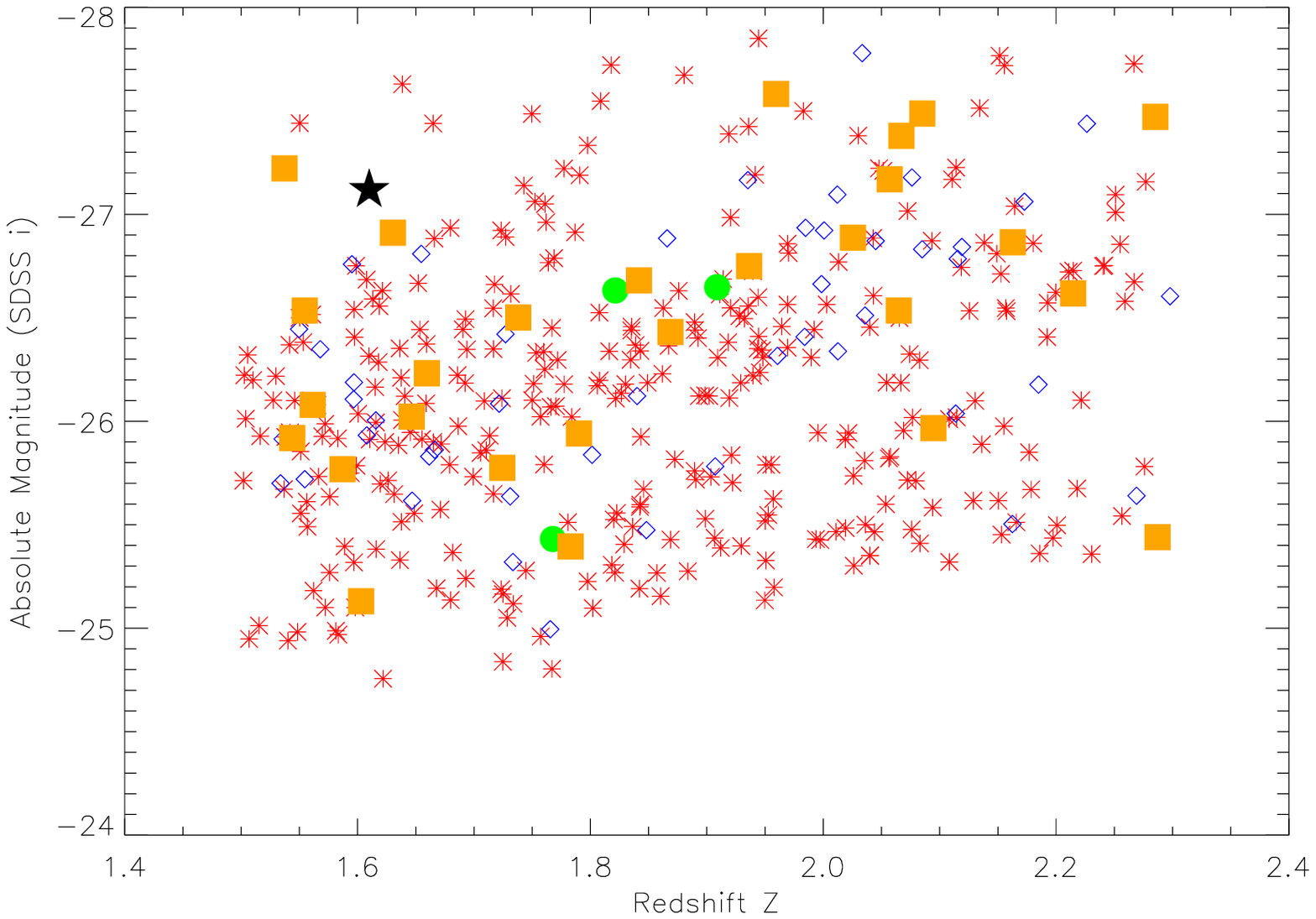} 
\label{fig:subabzselect}
}
\caption[dum]{Distributions of BAL and non-BAL QSOs as a function of redshift and absolute magnitude in SDSS $i$ band. The top panel shows the full sample before our selection methods have been used. The dashed box shows the redshift and absolute magnitude cut-offs used in this study. The lower panel shows the final sample after removal of those QSOs falling outside our absolute magnitude and redshift cutoffs, and after statistical matching (see text for details). Red asterisks are non-BAL QSOs, blue diamonds are HiBAL QSOs, the black filled star is our solitary LoBAL QSO, green circles are our detected HiBAL QSOs and orange squares are our detected non-BAL QSOs.}
\label{fig:selection}
\end{figure}

In this paper we use observations from the ESA \textit{Herschel} Space Observatory (\citealt{Herschdata}) as part of the \textit{Herschel} Astrophysical Terahertz Large-Area Survey (H-ATLAS, \citealt{Eales2010}) survey phase 1 data set, reduced in the same way as the SDP data set and with the same 5\,$\sigma$ flux limits (\citealt{Ibar2010, Pascale2011, Rigby2011}). It is the largest \textit{Herschel} open-time extragalactic key project on the observatory, and will eventually observe $\sim$550 sq deg in five passbands at 100, 160, 250, 350 and 500~$\mu$m with the PACS (\citealt{Poglitsch2010}) and the SPIRE (\citealt{Griffin2010}) instruments, allowing the study of interstellar dust at angular scales $\simeq$ 10 times smaller than with IRAS. With these data, approximately 250,000 galaxies out to redshifts of 3 to 4 are expected to be detected. We omit the PACS data (\citealt{Ibar2010}) from our analysis as the selected sources may possibly be contaminated by warmer dust heated by the AGN and the data do not reach the required depth to be useful. Instead, we have concentrated on the SPIRE data (\citealt{Pascale2011}). SPIRE wavelengths have been chosen because they are less likely to be contaminated by direct emission from the warm dust of the AGN torus, with a starburst component at these wavelengths being required to account for the total FIR emission, as suggested by the SED fits of \citet{Hatz2010}.

The \citet{Gib2009} catalogue has been cross-matched with the H-ATLAS 9, 12 and 15~h fields. Each field is approximately 12$^\circ$ in RA by 3$^\circ$ in Dec (6$^\circ$ by 3$^\circ$ for the 12~h field), which places a limit on the size of our final sample. Only those QSOs falling within the field boundaries are included resulting in an initial sample of 88. These further reduce to 83 due to 5 that have a C{\sc\,iv} Balnicity Index value of 0. These BAL QSOs have been classified using alternative lines such as Si{\sc\,iv} or Mg{\sc\,ii}, and are considered beyond the scope of this paper.

We next apply constraints within the parameter space $1.5 \leq z < 2.3$ and $-28 \leq  M_{\rm i} \leq -24$, where most of the BAL QSOs lie, in order to reduce the susceptibility of our study to the effects of luminosity or redshift evolution (e.g. \citealt{Bonfield2011}). This restriction gives two advantages: it allows easy construction of a non-BAL QSO comparison sample and limits the effects of cosmological evolution whilst comparing similar luminosities.
In this way the sample is further reduced to 50 BAL QSOs. We have also addressed the important distinction between the `classic' $BI$ and the `extended' version given in \citet{Gib2009}, $BI_{0}$, to see if this classification has any effect on the flux densities and other properties of our sample. As a result of this distinction, the total number of BAL QSOs following the `classic' $BI$ definition drops to 36. One of these, SDSS115404.13+001419.6, is a LoBAL QSO. This source is not considered in our stacking analyses in Section 4. The distribution of BAL QSOs over the 3 fields is as follows: 9~h field - 19, 12~h field - 12, 15~h field - 19.
Our comparison sample consists of 329 non-BAL QSOs drawn from the SDSS Data Release 5 \citet{Schneider2007}. We did not use the SDSS DR8 catalogue since there has been no corresponding BAL QSO catalogue produced. We trimmed objects at random from overpopulated parts of the comparison sample (determined via visual inspection of $M_{\rm i}$ and $z$) until a two-sample KS test comparison in $M_{\rm i}$ and $z$ between the non-BAL and BAL QSO sample returned a null--hypothesis probablility of $p$ $>$ 0.05; for the final non-BAL QSO comparison sample separate KS tests on redshift and $M_{\rm i}$ give $p$-values of 0.67 and 0.20 respectively (`extended sample') and 0.89 and 0.17 respectively (`classic' sample). Running a 2-d KS test (\citealt{Peacock1983}) on redshift and $M_{\rm i}$ for both samples returns a $p$-value of 0.27. We can therefore assume the populations are adequately matched in $M_{\rm i}$ and $z$. The lower panel of Fig.~\ref{fig:selection} shows the final BAL and non-BAL QSO samples on the ($z$, $M_{\rm i}$) plane while Table \ref{tab:BALdata} gives details of the final BAL QSO sample.

\begin{table*}
\scriptsize
\tabcolsep 3.5pt
\caption{The BAL QSO Sample. The classification method used to determine the Balnicity of a source is given in the Type/Detected column, along with whether it is detected. If a source is classified using BI, then it will also be classified as a BAL QSO by the $BI_{0}$ definition. The 250~$\mu$m flux density and signal-to--noise ratio are given derived from the 250~$\mu$m flux density divided by the sum in quadrature of the instrumental noise in the corresponding pixel and confusion noise in \citet{Rigby2011}. 5\,$\sigma$ detections found in the Phase 1 catalogue with matched optical/NIR counterparts (see Section 4.2) are shown with their flux density values taken from the Phase 1 catalogue. The average 5\,$\sigma$ limits at 250, 350 and 500~$\mu$m are 33.5, 37.7 and 44.0~mJy beam$^{-1}$. For undetected sources, their flux density values are the best estimates from measuring the flux density in the closest pixel of the PSF-smoothed map and noise map and are not reliably associated with R~$>$~0.80 to a 250~$\mu$m source as for our detections. The FIR data for the non-BAL Quasars within H-ATLAS will be dealt with in a future paper (Bonfield et al. in prep), however all maps and catalogues used in this paper will be released as part of the Phase 1 data release from the H-ATLAS website.}
\begin{centering}
\begin{tabular}{l c r  r  c  r  r  r  r  r  r}
\hline
Source & z & Type/ & C{\sc\,iv} EW & M$_{i}$ & 250~$\mu$m flux & SNR & 350~$\mu$m flux & SNR & 500~$\mu$m flux & SNR\\
& & Detected? & ($\rmn{\AA}$) & & density (mJy) & 250 & density (mJy) & 350 & density (mJy) & 500\\\hline
SDSSJ084307.36$-$001228.4 & 1.7271 & BI/No & 7.40 &  $-$26.421 & $+$18.43 & $+$2.82 & $+$13.44 & $+$1.86 & $+$6.00 & $+$0.75\\
SDSSJ084524.10$-$000915.4 & 2.0121 & BI/No & 56.10 & $-$27.095 & $+$18.38 & $+$3.18 & $+$2.79 & $+$0.41 & $+$2.67 & $+$7.84\\
SDSSJ084842.13+010044.3 & 1.6616 & BI/No & 20.00 & $-$25.830 & $-$4.13 & $-$0.63 & $-$3.79 & $-$0.52 & $-$4.00 & $-$0.047\\
SDSSJ085316.22+012052.0 & 1.6663 & BI/No & 19.80 & $-$25.862 & $+$13.58 & $+$2.11 & $+$12.29 & $+$1.68 & $+$3.83 & $+$0.44\\
SDSSJ085436.41+022023.5 & 1.9089 & BI/Yes & 20.70 & $-$26.648 & $+$55.87 & $+$8.24 & $+$63.14 & $+$7.90 & $+$39.87 & $+$4.51\\
SDSSJ085609.02+001357.7 & 1.8401 & BI/No & 6.50 &  $-$26.122 & $+$9.74 & $+$1.51 & $+$6.88 & $+$0.94 & $-$1.72 & $-$0.21\\
SDSSJ085647.99+003107.4 & 2.2979 & BI/No & 8.70 &  $-$26.604 & $+$21.94 & $+$3.45 & $+$15.61 & $+$2.15 & $+$20.58 & $+$2.39\\
SDSSJ090030.36+015154.9 & 1.9848 & BI/No & 10.00 & $-$26.935 & $-$4.41 &$-$0.67 & $-$3.42 & $-$0.46 & $-$0.786 & $-$0.09\\
SDSSJ090211.60+003859.5 & 1.5339 & BI/No & 7.60 &  $-$25.700 & $+$16.06 & $+$2.44 & $-$1.30 & $+$0.18 & $+$0.08 & $+$0.09\\  
SDSSJ090331.90+011804.5 & 1.9072 & BI/No & 17.60 & $-$25.782 & $+$11.88 & $+$1.85 & $+$3.48 & $+$0.47 & $+$10.01 & $+$1.18 \\
SDSSJ090517.24+013551.4 & 1.7678 & BI/Yes & 41.70  &$-$25.431 & $+$57.81 & $+$8.54 & $+$72.79 & $+$8.97 & $+$52.59 & $+$5.86\\
SDSSJ090523.07+001136.9 & 1.5600 & BI/No & 8.20 &  $-$26.348 & $+$22.02 & $+$3.44 & $+$19.48 & $+$2.74 & $+$7.55 & $+$0.90\\
SDSSJ090904.52$-$000234.5 & 1.7656 & BI$_{0}$/No & 8.50 &  $-$24.995 & $-$3.85 & $-$0.59 & $-$10.09 & $-$1.39 & $-$2.88 & $-$0.33\\
SDSSJ091110.29+004822.8 & 2.2691 & BI$_{0}$/No & 4.00 &  $-$25.640 & $+$3.85  & $+$0.66 & $+$4.19 & $+$0.63 & $+$2.94 & $+$0.38\\
SDSSJ091144.41+000423.6 & 1.8014 & BI$_{0}$/No & 5.70  & $-$25.838 & $+$3.08  & $+$0.53 & $+$0.23 & $+$0.03 & $+$4.79 & $+$0.62\\
SDSSJ091524.29+002032.6 & 1.9353 & BI$_{0}$/No & 6.00 &  $-$27.165 & $+$23.22 & $+$3.56 & $+$15.38 & $+$2.12 & $+$8.98 & $+$1.06\\
SDSSJ091600.60+011621.6 & 1.8481 & BI/No & 15.20 & $-$25.474 & $+$3.08 & $+$0.47 & $-$3.40 & $-$0.47 & $-$5.01 & $-$0.57\\
SDSSJ091808.80+005457.7 & 2.1155 & BI/No & 5.30 &  $-$26.785 & $+$3.73 & $+$0.56 & $-$0.65 & $-$0.08 & $-$3.37 & $-$0.39\\
SDSSJ091951.29+005854.9 & 2.1138 & BI/No & 6.90 & $-$26.037 & $+$4.62 & $+$0.70 & $+$7.60 & $+$1.04 & $+$9.19 & $+$1.05\\
SDSSJ113510.27$-$003558.2 & 1.7335 & BI/No & 8.30 &  $-$25.320 & $+$8.17 & $+$1.25 & $-$12.53 & $-$1.75 & $-$2.66 & $-$0.31\\ 
SDSSJ113537.56+004130.1 & 1.5498 & BI/No & 15.00 & $-$26.445 & $+$4.52 & $+$0.69 & $-$6.21 & $-$0.84 & $+$0.83 & $+$0.01\\
SDSSJ113544.33+001118.7 & 1.7311 & BI/No & 23.80 & $-$25.637 & $-$0.35 & $-$0.05 & $+$2.62 & $+$0.36 & $+$0.72 & $+$0.08\\
SDSSJ113651.54$-$002836.0 & 1.6157 & BI/No & 7.50  & $-$26.004 & $+$12.83 & $+$1.95 & $+$7.51 & $+$1.05 & $+$10.93 & $+$1.28\\
SDSSJ113934.63$-$005901.5 & 1.6079 & BI/No & 10.40 & $-$25.932 & $+$9.06 & $+$1.39 & $+$2.61 & $+$0.35 & $-$0.8 & $-$0.092\\
SDSSJ114259.29$-$000156.4 & 1.9840 & BI$_{0}$/No & 9.30 &  $-$26.408 & $-$3.91 & $-$0.59 &$-$2.86 & $-$0.39 & $+$11.61 & $+$1.38\\
SDSSJ114333.62+013709.0 & 1.5547 & BI/No & 22.70 & $-$25.720 & $+$21.98 & $+$3.36 & $+$29.35 & $+$3.97 & $-$4.46 & $-$0.53\\ 
SDSSJ114954.94+001255.3 & 1.5952 & BI/No & 4.90 &  $-$26.759 & $+$11.45 & $+$1.75 &$+$14.51 & $+$2.01 & $+$19.20 & $+$2.23\\
SDSSJ115404.13+001419.6 & 1.6100 & BI/No & 31.50 & $-$27.119 & $+$29.88 & $+$4.49 &$+$20.68 & $+$2.86 & $-$0.24 & $-$0.03\\
SDSSJ115407.74+001113.4 & 1.6547 & BI$_{0}$/No & 7.90 &  $-$26.809 & $+$32.72 & $+$4.94 & $+$38.02 & $+$4.47 & $+$38.02 & $+$4.46\\
SDSSJ115809.69$-$013754.3 & 1.5969 & BI/No & 17.70 & $-$26.188 & $-$0.84 & $-$0.13 & $+$4.61 & $+$0.64 & $-$4.81 & $-$0.57\\
SDSSJ115940.79$-$003203.5 & 2.0334 & BI$_{0}$/No & 5.70 &  $-$27.779 & $+$28.25 & $+$4.36 & $+$31.52 & $+$4.26 & $+$37.79 & $+$4.38\\
SDSSJ140842.75+010828.7 & 1.6469 & BI/No & 15.80 & $-$25.616 & $+$2.45 & $+$0.37 & $+$5.23 & $+$0.73 & $-$1.87 & $-$0.22\\
SDSSJ140918.72+004824.3 & 2.0008 & BI/No & 10.10 & $-$26.922 & $+$21.47 & $+$3.32 & $+$6.88 & $+$0.96 & $+$7.65 & $+$0.89\\
SDSSJ141434.39$-$011534.4 & 1.5967 & BI$_{0}$/No & 13.70 & $-$26.106 & $+$12.73 & $+$2.00 & $+$10.06 & $+$1.39 & $+$15.00 & $+$1.73\\
SDSSJ142050.33$-$002553.1 & 2.0850 & BI/No & 35.20 & $-$26.831 & $+$10.18 & $+$1.76 & $+$2.62 & $+$0.39 & $-$10.75 & $-$1.37\\
SDSSJ142423.76+001451.0 & 2.1849 & BI$_{0}$/No & 10.30 & $-$26.177 & $-$2.54 & $-$0.39 & $+$7.44 & $+$1.03 & $+$0.59 & $+$0.06\\
SDSSJ142820.59$-$005348.3 & 1.5357 & BI$_{0}$/No & 5.30 &  $-$25.914 & $+$23.56 & $+$3.63 & $+$24.28 & $+$3.37 & $-$5.39 & $-$0.64\\
SDSSJ143030.97+003440.1 & 1.9985 & BI$_{0}$/No & 14.40 & $-$26.663 & $+$9.03 & $+$1.56 & $+$3.67 & $+$0.56 & $-$3.08 & $-$0.40\\
SDSSJ143144.65+011644.1 & 1.9607 & BI/No & 20.50 & $-$26.316 & $+$6.50 & $+$1.02 & $+$10.24 & $+$1.41 & $+$5.51 & $+$0.66\\
SDSSJ143209.79+015256.3 & 2.1191 & BI/No & 23.20 & $-$26.842 & $+$16.09 & $+$2.51 & $+$12.22 & $+$1.69 & $+$6.43 & $+$0.76\\
SDSSJ143627.79+004655.7 & 2.1625 & BI$_{0}$/No & 7.90 & $-$25.503 & $+$16.87 & $+$2.64 & $+$1.24 & $+$0.17 & $+$2.83 & $+$0.33\\
SDSSJ143641.24+001558.9 & 1.8659 & BI/No & 20.80 & $-$26.884 & $+$3.13 & $+$0.49 & $-$7.55 & $-$1.03 & $-$11.05 & $-$1.27\\
SDSSJ143758.06+011119.5 & 2.0450 & BI/No & 18.80 & $-$26.871 & $+$3.56 & $+$0.59 & $+$12.90 & $+$1.89 & $+$5.59 & $+$0.69\\
SDSSJ143907.51$-$010616.7 & 1.8214 & BI$_{0}$/Yes & 7.60 &  $-$26.632 & $+$73.35 & $+$10.55 & $+$42.41 & $+$5.34 & $+$29.32 & $+$3.31\\
SDSSJ144256.86$-$004501.0 & 2.2264 & BI/No & 2.50 & $-$27.438 & $+$9.88 & $+$1.71 & $+$19.66 & $+$2.96 & $+$18.44 & $+$2.36\\
SDSSJ144434.80+003305.3 & 2.0359 & BI$_{0}$/No & 7.60 &  $-$26.511 & $+$8.44 & $+$1.43 & $-$2.70 & $-$0.40 & $-$5.33 & $-$ 0.67\\
SDSSJ144911.82$-$010014.8 & 2.1728 & BI/No & 39.60 & $-$27.061 & $+$4.12 & $+$0.64 & $+$0.29 & $+$0.04 & $+$3.75 & $+$0.49\\
SDSSJ144959.96+003225.3 & 1.7217 & BI/No & 11.00 & $-$26.085 & $+$11.90 & $+$1.87 & $+$1.18 & $+$0.16 & $-$4.06 & $-$0.48\\
SDSSJ145045.42$-$004400.3 & 2.0762 & BI/No & 18.00 & $-$27.178 & $+$10.38 & $+$1.60 & $-$1.79 & $-$0.24 & $-$0.68 & $-$0.08\\
SDSSJ145511.44+002146.0 & 2.0126 & BI/No & 15.80 & $-$26.338 & $+$0.61 & $+$0.09 & $-$2.44 & $-$0.34 & $+$3.80 & $+$0.45\\
\hline
\end{tabular}
\label{tab:BALdata}
\end{centering}
\end{table*}

In the next section, we compare the FIR properties of matched BAL and non-BAL QSO samples, and determine SFRs for the samples based on separate classification schemes (as explained in Section 2). Our study improves on previous work since whilst the H-ATLAS survey is still not complete, the three fields in the H-ATLAS phase 1 dataset centred at 9, 12, and 15h give us a large area providing a uniformly selected sample of objects cross-matched with the Sloan Digital Sky Survey (SDSS).

\section{Method and results}

Here we determine the FIR properties of the selected QSOs and investigate how the BAL QSOs relate to the non-BAL QSOs. We further  discuss the effects of different selection criteria  i.e., $BI$ versus $BI_{0}$. 
  
\begin{figure*}
\begin{center}$
\begin{array}{ccc}
\begin{overpic}[width=1.25in]{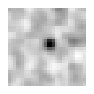}
  \put(3,19){\scriptsize \color{red} 250 $\mu$m} 
  \put(3,12){\scriptsize \color{red} `Extended'}
  \put(3,5){\scriptsize \color{red} 49 sources}
  \end{overpic} &
\begin{overpic}[width=1.25in]{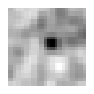}
  \put(3,19){\scriptsize \color{red} 350 $\mu$m} 
  \put(3,12){\scriptsize \color{red} `Extended'}
  \put(3,5){\scriptsize \color{red} 49 sources}
  \end{overpic}  &
\begin{overpic}[width=1.25in]{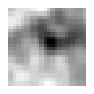}
  \put(3,19){\scriptsize \color{red} 500 $\mu$m} 
  \put(3,12){\scriptsize \color{red} `Extended'}
  \put(3,5){\scriptsize \color{red} 49 sources}
  \end{overpic}  \\
\begin{overpic}[width=1.25in]{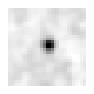}
  \put(3,19){\scriptsize \color{red} 250 $\mu$m} 
  \put(3,12){\scriptsize \color{red} Non-BAL}
  \put(3,5){\scriptsize \color{red} 329 sources}
  \end{overpic} &
\begin{overpic}[width=1.25in]{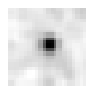}
  \put(3,19){\scriptsize \color{red} 350 $\mu$m} 
  \put(3,12){\scriptsize \color{red} Non-BAL}
  \put(3,5){\scriptsize \color{red} 329 sources}
  \end{overpic} &
\begin{overpic}[width=1.25in]{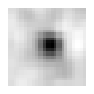}
  \put(3,19){\scriptsize \color{red} 500 $\mu$m} 
  \put(3,12){\scriptsize \color{red} Non-BAL}
  \put(3,5){\scriptsize \color{red} 329 sources}
  \end{overpic}  \\
\end{array}$
\end{center}
\caption[dum]{The 250, 350 and 500~$\mu$m weighted mean flux density stacks of the `extended' BAL and non-BAL QSOs (top and bottom row respectively) samples. Each postage stamp image is 210$\times$210 arcsec$^{2}$. In each corner in red is given the wavelength, classification, and the number of sources used to create that stack.}
\label{fig:balstacks}
\end{figure*}

\subsection{Stacking analysis}

The FIR flux densities used in this section are taken directly from the final PSF-convolved images for all three H-ATLAS fields. 
Cutouts of set size (210$\times$210 arcsec$^{2}$) of the region surrounding each BAL/non-BAL QSO are generated, and these are co-added, each pixel being the weighted mean flux density in that pixel.
The weighted mean stacked flux density value of each source is taken to be that at the centre of the stacked image which is the closest pixel to the catalogue position in either the \citet{Gib2009} BAL or \citet{Schneider2007} SDSS QSO samples. We background subtract by randomly selecting a sample of 40,000 pixels from each field to create a mean background for that field at that wavelength. We then subtract this mean background from every pixel in our cutout images for each QSO. These mean values for each wavelength are shown in Table~\ref{tab:BALtable}, while Fig.~\ref{fig:balstacks} shows the BAL and non-BAL QSO stacked images at each wavelength. 

We find that at longer wavelengths the stacked BAL QSO image is less distinguishable from the background, which is expected due to increased confusion noise and a higher instrumental noise; galaxies at 500~$\mu$m are also expected to be fainter, since the Rayleigh--Jeans tail is being sampled. Table~\ref{tab:BALtable} shows the central pixel values of the stacks. We find that there is no statistical difference between BAL QSOs and non-BAL QSOs; their flux densities in all bandpasses are the same within the errors. This result holds for both the `classic' and `extended' BAL QSO samples. 

We attempt to reproduce our results using the Institut d'Astrophysique Spatiale (IAS) stacking library. It stacks data to allow a statistical detection of a faint signal using positions of galaxies detected at shorter wavelengths as with our method but has also been tested and validated with galaxy clustering in mind, something which may affect our results (see \citealt{Bavouzet2008, Bethermin2010} for further details). However, we still find no difference between the BAL QSO and non-BAL QSO samples in each bandpass and the IAS values are in agreement with our values within the errors (see Table \ref{tab:BALtable}). 

It should be noted that submillimetre fluxes may still be overestimated due to clustering of sources if they emit in the SPIRE bands, but since we cannot identify the sources around the QSO at high redshifts there is little that can be done to solve this problem. We also emphasize that the effects of confusion should be the same for both BAL QSOs and non-BAL QSOs since they are treated in an equal manner within both methods.

\begin{table*}
\scriptsize
\tabcolsep 2.5pt
\caption{The BAL (`extended' and `classic' samples) and non-BAL QSO FIR weighted mean flux densities in the 250, 350 and 500~$\mu$m bandpasses. The number in brackets gives the number of objects within each stack. We also include here our determined background noise for each field at each wavelength.}
\begin{centering}
\begin{tabular}{c  c  c	c	c	c	c	c	c	c}
\hline
 & \multicolumn{3}{c}{BAL QSO and non-BAL QSO flux densities} & \multicolumn{3}{c}{IAS BAL QSO and non-BAL QSO flux densities}  & \multicolumn{3}{c}{Mean background in each field} \\ \hline
Bandpass & `Extended' stack & `Classic' stack & non-BAL stack & `Extended' stack & `Classic' stack & non-BAL stack & 9~h field & 12~h field & 15~h field \\
($\mu$m) & (49, mJy) & (35, mJy) & (329, mJy) & (49, mJy) & (35, mJy) & (329, mJy) & (mJy) & (mJy) & (mJy)\\ \hline 
250 & 11.81$\pm$1.20 & 10.97$\pm$1.51 & 10.94$\pm$0.53 & 12.34$\pm$1.41 & 10.87$\pm$1.61 & 11.74$\pm$0.63 & 1.14$\pm$0.0003 & 0.91$\pm$0.0003 & 1.45$\pm$0.0004 \\ 
350 & 8.76$\pm$1.52 & 8.13$\pm$2.11 & 8.97$\pm$0.42 & 10.80$\pm$1.61 & 8.79$\pm$1.79 & 9.57$\pm$0.69 & 2.82$\pm$0.0003 & 2.72$\pm$0.0003 & 3.05$\pm$0.0003\\ 
500 & 5.78$\pm$1.66 & 4.55$\pm$2.11 & 7.25$\pm$0.54 & 7.72$\pm$1.41 & 5.48$\pm$1.86 & 7.85$\pm$0.62 & 0.30$\pm$0.0003 & 0.27$\pm$0.0003 & 0.84$\pm$0.0003\\
\hline
\end{tabular}
\label{tab:BALtable}
\end{centering}
\end{table*}

\begin{figure}
\centering
\begin{overpic}[scale=0.5]{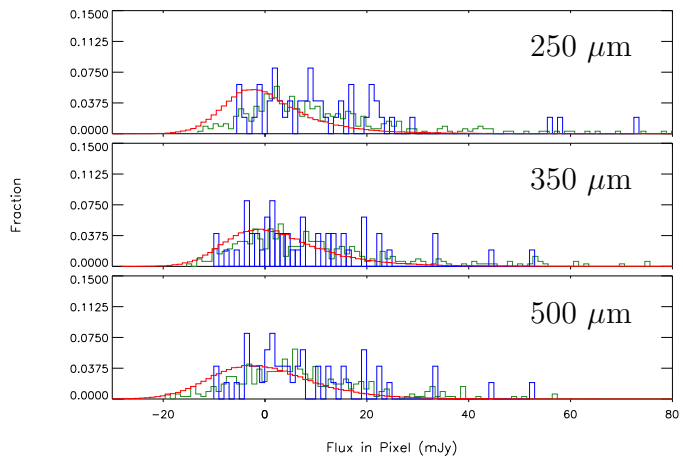}
  \put(75,60){\Large \color{black} 250 $\mu$m} 
  \put(75,40){\Large \color{black} 350 $\mu$m}
  \put(75,20){\Large \color{black} 500 $\mu$m}
  \end{overpic}
\caption{The distribution of flux density at each wavelength for the full BAL (blue outline) and non-BAL QSO (green outline) samples in bins of width 1~mJy. The random background flux densities are shown as a red outline.} 
\label{fig:sigevidence}
\end{figure}

The Gaussian errors quoted in Table~\ref{tab:BALtable} must be treated with caution, since background noise will be non-Gaussian owing to confusion noise. To determine quantitatively whether the stacked BAL QSOs are detected significantly, flux densities from 120,000 randomly chosen positions in the field have been measured following \citet{Hardcastle2010}. The random background compared with the BAL and non--BAL QSO fluxes are shown in Figure \ref{fig:sigevidence}. Using a KS test, we can examine whether the flux densities of the stacks are statistically distinguishable from those taken from randomly chosen positions, as a KS test is not influenced by the noise properties.
\begin{table*}
\scriptsize
\tabcolsep 5.8pt
\caption{The KS statistics and probabilities of each sample being indistinguishable from a randomly selected sample of flux densities taken from the H-ATLAS fields. The fraction shows how many random stacks had flux densities greater than our sample stacks. }
\begin{centering}
\begin{tabular}{c  c  c	c	c	c	c	c	c	c}
\hline
& \multicolumn{3}{c}{'Extended' $BI_0$} & \multicolumn{3}{c}{`Classic' $BI$} & \multicolumn{3}{c}{non-BAL QSOs} \\ \hline
Bandpass ($\mu$m) & KS statistic & KS probability & Fraction & KS statistic & KS probability & Fraction & KS statistic & KS probability & Fraction\\ \hline
250 & 0.48 & $< 10^{-10}$ & $< 10^{-5}$ & 0.48 & $< 10^{-10}$ & 0.00013 & 0.38 & $< 10^{-10}$ & $< 10^{-5}$\\ 
350 & 0.32 & $< 10^{-10}$ & 0.00068 & 0.33 & $< 10^{-10}$ & 0.0028 & 0.29 & $< 10^{-10}$ & $< 10^{-5}$ \\
500 & 0.28 & 0.00051 & 0.00048 & 0.29 & 0.0041 & 0.022 & 0.25 & $< 10^{-10}$ & $< 10^{-5}$ \\ 
\hline
\end{tabular}
\label{tab:significance}
\end{centering}
\end{table*}

In comparison to the background we detect the BAL QSOs at all wavelengths using both classification schemes. The significance of the 250~$\mu$m detection for `extended' BI$_{0}$ BAL QSOs is well over 5\,$\sigma$ ($p < 10^{-10}$), and as expected, the significance decreases towards longer wavelengths, being lowest at 500~$\mu$m ($p \sim 5\times10^{-4}$). The larger sample of non-BAL QSOs is also detected with high significance above the background at all wavelengths ($p < 10^{-10}$). Returned KS probabilities are shown in Table~\ref{tab:significance}.
We also used a KS test to compare the distribution of flux densities between the BAL QSO samples and the non-BAL QSO samples (see Fig. \ref{tab:significance}). At 250, 350 and 500~$\mu$m their $p$-values are $p=0.39$, $0.80$, and $0.37$ respectively for the `extended' sample when comparing the BAL QSO sample with the full non-BAL QSO sample. We therefore cannot reject the null hypothesis that the BAL and non--BAL QSOs are drawn from the same underlying flux density distibution.

We performed a second test where the 120,000 stacks were created from flux densities extracted from random positions, with the number of elements in the stack equal to the elements in the comparison sample stack, i.e. 49, 35 and 329, and we then compared these random stacks with the stacked flux densities of our BAL and non-BAL QSO samples. The fraction of random stacks where the weighted mean flux density exceeds the weighted mean flux density in our sample stacks provides an estimate of the probability of the detection of our stacks being just a background fluctuation. We find that 250~$\mu$m is generally of greater significance than that at 500~$\mu$m for the BAL QSO samples (e.g. $p< 10^{-5}$, $p=4.8\times10^{-4}$ respectively for the `extended' BI$_{0}$ BAL QSOs in each bandpass), and that none of the 120,000 flux density stacks extracted at random positions have larger weighted mean flux densities than the BAL and non-BAL stacked flux density as shown in Table \ref{tab:significance}. This indicates that our weighted mean flux density values in Table \ref{tab:BALtable} are not due to background fluctuations.

\subsection{Detection rates}

We now investigate how many individual BAL QSOs are significantly detected in the H--ATLAS data at the $>$5\,$\sigma$ level.
The Phase 1 H-ATLAS catalogue used a likelihood ratio technique to identify optical and near-infrared counterparts with objects detected by \textit{Herschel} as with the SDP catalogue (\citealt{Smith2010}) in the phase 1 data. We find that three BAL QSOs from our sample are associated with (5\,$\sigma$) 250\,$\mu$m sources in the $H$-ATLAS Phase 1 catalogue (i.e. they have reliability, $R > 0.80$); two of these are "classic" $BI$-defined BAL QSOs. For the non-BAL QSOs, 27 of the 329 have reliable $R>0.80$ 250\,$\mu$m counterparts. This raises the question of whether there is an actual difference in the detection rate of BAL QSOs and non-BAL QSOs. We detect 6.0 per cent of the `extended' sample (8.3 per cent of the `classic' $BI$ defined sample) compared to 8.2 per cent of the non-BAL sample.  Comparing these detection rates using a binomial probability distribution gives a null result ($p=0.19$ and $p=0.23$ for the `extended' and `classic' sample respectively). We cannot therefore reject the null hypothesis that the detected HiBALs are drawn from the same distribution as the detected non-BAL QSOs, consistent with our conclusions above on their flux density distribution.

\subsection{FIR luminosities and star-formation rates}

We have limited information on the shape of the SED of our QSOs and therefore must choose a suitable template in order to compute FIR luminosities ($L_{\rm FIR}$). The primary galaxy that we have chosen for this purpose is Mrk~231. Of the local ULIRGs, it is the most luminous, with a rest--frame luminosity at $8-1000\ \mu$m of $3.2 \times 10^{12}$~L$_{\odot}$ (\citealt{Sanders2003}), and it shows similar broad absorption line features to our sample, having been classified as a BAL QSO, strictly a LoBAL QSO (\citealt{Smith1995}). We have also investigated the effect of using a template based on IZw1, a local Seyfert type 1 galaxy in the nearby universe at $z = 0.0589$ (\citealt{Ho2009}). We assume $T=45$~K and emissivity index $\beta=1.6$ for Mrk 231, and $T=36$~K and $\beta=1.7$ for IZw1 from fitting a greybody to data points retrieved from the NASA Extragalactic Database (32 and 9 respectively). 


We determine the FIR luminosity via a ratio method between the 250~$\mu$m flux density of our greybody template if placed at the redshift of the QSO in our sample and the 250~$\mu$m flux density of the QSO. This implicitly assumes that the QSO has the same FIR SED as the greybody template.
The calculation of the SFR was then performed in the standard manner using the relationship with $L_{\rm FIR}$ published by \cite{Kennicutt1998} in equation \ref{eq:sfr} and these are presented in Table \ref{tab:SFRtable}, i.e.
\begin{equation}\label{eq:sfr}
SFR(M_{\odot}yr^{-1}) = 4.5 \times 10^{-44} L_{FIR}~~(ergs^{-1}).
\end{equation}
The \cite{Kennicutt1998} relation is dependent on the models of \citet{LeithHeck1995}. These models were specifically created to correspond to the conditions that are prevalent within giant H{\sc ii} regions, H{\sc ii} galaxies, nuclear starbursts and infrared luminous starburst galaxies and traces diffuse gas within these regions. The \cite{Kennicutt1998} relation assumes continuous starbursts of age $10-100$~Myr, and requires the integrated IR luminosity over the range $8-1000\ \mu$m.  

\begin{figure}
\centering
\includegraphics[scale=0.5]{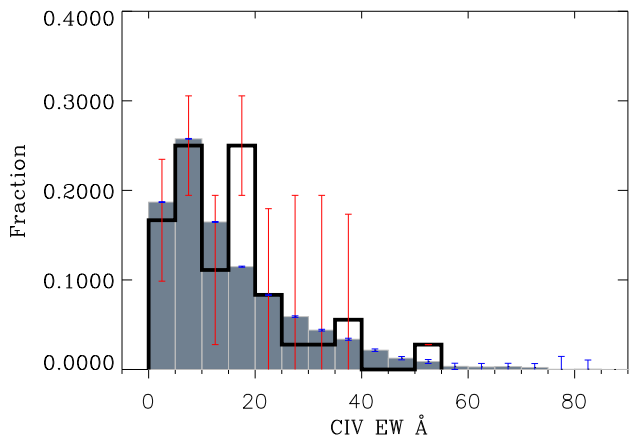} 
\caption{The distribution of C{\sc\,iv} absorption--line equivalent width for the full \citet{Gib2009} sample (grey) and our `classic' BAL QSO sample (black outline) in bins of  width 5\,\AA. One source in the \citet{Gib2009} sample is omitted here, SDSS090901.80+535935.7, having a C{\sc\,iv} absorption--line equivalent width of 193.9\,\AA. All others have C{\sc\,iv} absorption--line equivalent widths below 90\,\AA.} 
\label{fig:GibCIV}
\end{figure}


\begin{table*}
\scriptsize
\tabcolsep 2pt
\caption{The star-formation rates and FIR luminosities of the detected (marked with an asterisk) BAL QSOs within the sample, along with the average star-formation rate in the 250, 350 and 500~$\mu$m bandpasses and weighted mean for the `extended' , `classic' and non-BAL samples as a whole using Mrk~231 and IZw1 as templates. We also include here our rare LoBAL QSO 115404.13+001419.6. We emphasize however that it is not reliably associated with R~$>$~0.80 to a 250~$\mu$m source as with our detections and instead is the best estimate from measuring the flux density in the closest pixel of the PSF-smoothed map. At 500~$\mu$m it has a negative flux so SFRs are not included. The quoted $L_{\rm FIR}$ (using our ratio method for 250~$\mu$m) and SFR estimates are determined from weighted stacking of the individual luminosities and SFRs of each galaxy rather than calculated from the weighted stack values in Table \ref{tab:BALtable} (shown in the flux density column), though the error will be underestimated as a result of confusion noise.}
\begin{centering}
\begin{tabular}{l  c  c  c  c c c c c c c c c} 

\hline
SDSS Source & z & Flux density & \multicolumn{2}{c}{$\log_{10}(L_{\rm FIR} / L_{\odot})$} & \multicolumn{8}{c}{SFR (M$_{\odot}$\,yr$^{-1}$)}\\
& & (250~$\mu$m) & (Mrk~231) & (IZw1)& \multicolumn{4}{c}{(Mrk~231)} & \multicolumn{4}{c}{(IZw1)}\\ 
& & (mJy)  & (250~$\mu$m) & (250~$\mu$m) & 250 & 350 & 500 & Mean & 250 & 350 & 500 & Mean\\\hline
*085436.41+022023.5 & 1.9089 & 55.87$\pm$6.81 & 12.87$\pm$0.07 & 12.75$\pm$0.10 & 1322$\pm$199 & 2353$\pm$364 & 3158$\pm$754 & 1559$\pm$175 & 1007$\pm$227 & 1506$\pm$343 & 1824$\pm$532 & 1159$\pm$189\\

*090517.24+013551.4 & 1.7678 & 57.81$\pm$6.77 & 12.85$\pm$0.06 & 12.71$\pm$0.10 & 1248$\pm$183 & 2386$\pm$369 & 4092$\pm$787 & 1473$\pm$164 & 921$\pm$204 & 1625$\pm$357 & 2341$\pm$597 & 1094$\pm$177\\

*143907.51$-$010616.7 & 1.8214 & 73.35$\pm$6.95 & 12.97$\pm$0.06 & 12.84$\pm$0.09 & 1640$\pm$212 & 1535$\pm$318 & 2297$\pm$724 & 1608$\pm$176 & 1224$\pm$259 & 972$\pm$269 & 1320$\pm$471 & 1103$\pm$187\\
115404.13+001419.6& 1.6100 & 29.88$\pm$6.65 & 12.51$\pm$0.10 & 12.37$\pm$0.13 & 580$\pm$139 & 694$\pm$250 & \dots & 607$\pm$121 & 413$\pm$121 & 427$\pm$170 & \ldots & 418$\pm$99\\
`Extended' BAL QSO & \ldots & 11.81$\pm$1.20 & 12.13$\pm$0.04 & 11.92$\pm$0.05 & 240$\pm$21 & 297$\pm$38 & 429$\pm$94 & 253$\pm$18 & 149$\pm$17 & 151$\pm$25 & 199$\pm$56 & 150$\pm$14\\
`Classic' BAL QSO & \ldots & 10.97$\pm$1.51 & 12.10$\pm$0.05 & 11.90$\pm$0.06 & 223$\pm$25 & 271$\pm$44 & 336$\pm$111 & 235$\pm$22 & 141$\pm$19 & 138$\pm$29 & 154$\pm$65 & 140$\pm$16\\
Non-BAL QSO & \ldots & 10.94$\pm$0.53 & 12.08$\pm$0.02 & 11.84$\pm$0.02 & 212$\pm$8 & 293$\pm$15 & 528$\pm$36 & 230$\pm$7 & 124$\pm$6 & 148$\pm$10 & 261$\pm$22 & 130$\pm$5\\
\hline
\end{tabular}
\label{tab:SFRtable}
\end{centering}
\end{table*}


The average Kennicutt SFR is $240\pm21$~M$_{\odot}$\,yr$^{-1}$ for the `extended' and $223\pm25$~M$_{\odot}$\,yr$^{-1}$ for the `classic' sample (Table~\ref{tab:SFRtable}); these values are consistent. Note that these estimates are determined from weighted stacking of the individual luminosities and SFRs derived from the 250~$\mu$m flux density of each galaxy rather than calculated from the weighted stacked flux density values in Table \ref{tab:BALtable}. The BAL and non-BAL average SFRs are also consistent within the errors. We find that the 3 detected BAL QSOs have ${\rm SFR} >1000$~M$_{\odot}$\,yr$^{-1}$. Since we cannot exclude the possibility that some fraction of the FIR emission comes from dust heated by the AGN, these SFRs should be regarded as upper limits. We also find that the average FIR luminosities presented in Table \ref{tab:SFRtable} for both the BAL and non-BAL QSO samples are, on their own, sufficiently large to classifiy them as ULIRGs. We note that the derived SFRs for each SPIRE band for IZw1 are more consistent than those determined for Mrk 231. This suggests that dust in BAL QSOs at higher $z$ is likely cooler than that found in the low-$z$ analogue Mrk 231.
Finally, we have also investigated the effect of varying the dust temperature and $\beta$ values and find that for plausible values ($25~\rm{K} \leq T \leq60$~K and $1.1\leq\beta\leq1.8$) there is about a factor of 2 variation in the final derived SFR values which does not change our conclusions.


\subsection{FIR luminosity and C{\sc\,iv} absorption--line equivalent width}
From an analysis of submillimetre wavelength data, \citet{Priddey2007} found a weak correlation between FIR flux density and C{\sc\,iv} absorption--line equivalent width and a rather stronger statistical link between submillimetre detection rate and C{\sc\,iv} absorption--line equivalent width; i.e., of the 15 sources they observed, all 6 ($>$2\,$\sigma$) detections in the 850~$\mu$m band had a C{\sc\,iv} absorption--line equivalent width of $\geq 25$\,\AA. A KS test returned a probability of 0.01 that the detections and non-detections had the same C{\sc\,iv} absorption--line equivalent width.

The small number of detections in our sample means we cannot perform the same tests that Priddey et al. used. In our sample there are only 5 sources with C{\sc\,iv} absorption--line equivalent widths of $\ga 25$\,\AA~in the 250~$\mu$m bandpass, of which 3 are detections. Assuming a binomial probability distribution we estimate the probability of detections having C{\sc\,iv} absorption--line equivalent widths $\ga 25$\,\AA~by chance is only 0.005. However, these are detections in one bandpass alone and the number of objects with C{\sc\,iv} absorption--line equivalent width $\ga 25$\,\AA~is very small. We must therefore choose a different C{\sc\,iv} absorption--line equivalent width cut-off for our HiBAL QSOs, such that we have similar sized samples to compare. We set this cut-off at $\geq 20$\,\AA, which gives us 11 HiBAL QSOs to compare with the remaining 24 in the classic sample. For completeness we tested whether the distribution of C{\sc\,iv} absorption--line equivalent widths within our restricted BAL QSO sample was representative of those within the parent population as described by \citet{Gib2009}. A KS test on the C{\sc\,iv} absorption--line equivalent widths of each population returns $p=0.12$, so we are able to assume our sample is representative. Figure~\ref{fig:GibCIV} shows histograms of the two C{\sc\,iv} absorption--line equivalent width distributions. For this analysis we studied the `classic' BAL QSO sample only since it suffers from less contamination from mis--classified sources than the `extended' sample due to the strictness of the BI mechanism. Furthermore, any true underlying physical relationship involving C{\sc\,iv} absorption--line equivalent width will be expected to correlate with luminosity rather than flux density. We therefore determined the $L_{\rm FIR}$ luminosities of each BAL QSO using our assumed T and $\beta$ for Mrk 231 and IZw1, separated them using our 20\,\AA~cutoff and then performed a Mann-Whitney test (\citealt{Mann1947}) to determine whether a greater luminosity correlates with a larger C{\sc\,iv} absorption--line equivalent width. For the Mrk 231 and IZw1 templates respectively, the test returned probabilities of 0.23 and 0.22, which indicates that there is no significant evidence for a difference in the two subsamples. Obviously with this method there is the caveat that the BAL QSOs must each have a similar SED shape to our greybody templates but these results indicate there is no evidence for a dependence of FIR luminosity on C{\sc iv} EW. 

\section{Discussion}

Our study shows that the FIR properties of HiBAL QSOs are statistically indistinguishable from those of non-BAL QSOs. We also find that FIR emission has no dependence on C{\sc\,iv} absorption--line equivalent width. Therefore the FIR luminosity has no dependence on the absorption strength of the outflow, and our results do not require an evolutionary link between HiBAL and non-BAL QSOs. Rather a simple orientation effect argument is sufficient. 

\citet{Priddey2007} found tentative evidence for a link between C{\sc\,iv} absorption--line equivalent width and submillimetre wavelength detection (albeit with a 2\,$\sigma$ detection threshold). As discussed by Priddey et al. such a link could be used to argue that BAL QSOs are in a distinct evolutionary phase along the lines proposed by \citet{Page2004,Page2011} where the QSO wind is in the process of terminating an epoch of enhanced star formation. Our results do not support this weak evidence for such a link between FIR output and C{\sc\,iv} absorption--line equivalent width. The discepancy may be linked to a possible selection effect discussed by Priddey et al.; that the most reddened QSOs must be intrinsically more luminous to meet their blue selection criterion, and therefore the most extreme objects, which might host the most powerful outflows would have higher dust reddening and higher FIR/submm output.
This could explain why we see no such correlation, since the SDSS $i$ band does not suffer reddening to as large a degree as the bluer bands in which past samples of QSOs were selected. 

However, a caveat to our results is that our sample is composed almost entirely of HiBAL QSOs. We have not considered either LoBAL QSOs or their rarer FeLoBAL cousins, which may well show FIR properties inconsistent with a simple orientation scheme hypothesis (see below). We note that the samples of \citet{Willott2003} and \citet{Priddey2007} were also mostly composed of HiBAL QSOs. The \citet{Priddey2007} sample contains only 1 known LoBAL QSO, and many of the \citet{Willott2003} QSOs classified as LoBAL QSOs are uncertain due to Mg{\sc\,ii} moving out of the spectral range at $z>2.26$. 

Regarding studies at other wavelengths, the findings of \citet{Green2001} carried out using the \textit{Chandra} X-ray observatory provide weak evidence that HiBAL QSOs appear X-ray weak due to intrinsic absorption, and that their underlying emission is actually the same as that of non-BAL QSOs (\citealt{Gallagher2002, Gallagher2006, Gibson2009a}), which supports an orientation scheme. In contrast, even taking into account intrinsic absorption, LoBAL QSOs remain X-ray weak. An analysis of X-ray absorption performed by \citet{Streb2010} suggests that LoBALs and HiBALs may be physically different objects. This is reinforced by composite SEDs ranging from X-ray to radio generated by \citet{Gallagher2007} who find similarities between the SEDs of HiBALs and non-BAL QSO but differences between these populations and LoBALs. However, it should be noted that the absorbed nature of BAL QSOs means that their X-ray data are limited in quality, so these findings cannot be taken as conclusive.
 
Certainly LoBAL QSOs are found to have redder spectra than other types (\citealt{Reich2003}), implying a larger dust mass in the host galaxy and/or vicinity of the SMBH. The idea that LoBALs and FeLoBALs form a separate evolutionary class finds support from the work of \citet{Farrah2007} and \citet{Urrutia2009}. \citet{Farrah2007} argue that FeLoBAL QSOs are galaxies where a massive merger driven starburst is ending, and the last remnants of a dust cocoon are being removed by a rapidly accreting supermassive black hole at the galaxy's centre. The SED fits of those detected at longer wavelengths (4/9) require a starburst component of the order of several hundred solar masses per year, which is taken to imply that FeLoBAL QSOs are associated with ULIRGs. Similar conclusions are drawn in \citet{Farrah2010} who find that FeLoBAL QSOs span a range of spectral shapes, consistent with the idea that they may be a transition phase in the lifetime of an AGN. However, as noted in their papers, the small and inhomogeneous sample sizes render such conclusions tentative. A larger sample was presented by \citet{Farrah2012} of 31 objects with optical to far--infrared photometry. These were found to all be highly luminous ($>10^{12}$~L$_{\odot}$), yet the bulk of IR emission in the majority of sources came from the AGN rather than a starburst component. The mid--IR and far--infrared properties of the FeLoBAL QSOs spanning the redshift range $0.8<z<1.8$ presented by Farrah et al. further reinforce the idea that they are a class apart from the general QSO population.

\citet{Urrutia2009} also found an unusually high percentage (32 per cent using the conservative $BI$ selection) of LoBAL QSOs in a sample of dust reddened type-1 QSOs; in fact all of these objects can also be classified as FeLoBALs.
The orientation hypothesis is irreconcilable with such findings allowing an evolutionary hypothesis to be invoked, even with the caveats offered by \citet{Urrutia2009}. 


The subset of the full \citet{Gib2009} sample studied here contains only one QSO, SDSS$115404.13+001419.6$, classified as a LoBAL. These objects are rare, making up only 15 per cent of all observed BAL QSOs. A very deep C{\sc\,iv} absorption trough (greater than anything seen in the HiBAL sample), combined with a high flux density (29.88~$\pm$~6.65~mJy at 250~$\mu$m) makes it a conspicous member of our sample. However, little can be said of an entire population from one object. Future work in the FIR/submillimetre wavebands targeting LoBALs and/or FeLoBALs would be invaluable.
 
\section{Conclusions}

\begin{enumerate}

\item Using a stacking analysis we have determined that HiBAL QSOs at $1.5\leq z < 2.3$ are statistically indistinguishable in terms of FIR luminosity (and therefore SFR) from a matched sample of non-BAL QSOs. This result is broadly consistent with previous work conducted at submillimetre wavelengths  (\citealt{Willott2003, Priddey2007}). The average FIR luminosities of both our HiBAL and non-BAL samples are $\ga 10^{12}$~L$_{\odot}$, sufficient to classify them as ULIRGs. We calculate weighted mean SFRs (strictly upper limits using this relation due to the possibility of dust heating by the AGN and our BAL QSOs having overdensities of neighbours) for the HiBALs of $240\pm21$~M$_{\odot}$\,yr$^{-1}$ for the `extended' and $223\pm25$~M$_{\odot}$\,yr$^{-1}$ for the `classic' sample from the flux density at 250~$\mu$m.

\item While \citet{Priddey2007} found tentative evidence for a dependence of submillimetre flux density on C{\sc\,iv} absorption--line equivalent width, we find no such dependence at FIR wavelengths. We suggest that the Priddey et al. result may have been due to small number statistics and selection effects as noted in the discussion.

\item Within our samples, 3/49 HiBALs and 27/329 non-BALs are detected at $>5$\,$\sigma$ significance. The detection rates for the two species are statistically indistinguishable.

\item Taken together these results suggest that BALs (strictly HiBALs) can be unified with non-BAL QSOs within a simple orientation scheme where a BAL QSO is observed only if a nuclear outflow intercepts our line-of-sight. However, with the current data we are unable to say whether (Fe)LoBALs can similarly be accommodated within this scheme or whether they form a distinct population perhaps caught at a key phase in their evolution. Future observations at FIR/submillimetre wavelengths would be valuable in this respect.

\end{enumerate}

\section*{Acknowledgements}

J.M.C.O. would like to thank STFC for a studentship and we thank the anonymous referee for useful comments that have improved the paper.
The Herschel-ATLAS is a project with Herschel, which is an ESA space observatory with science instruments provided by European-led Principal Investigator consortia and with important participation from NASA. The H-ATLAS website is http://www.h-atlas.org/.

This research has made use of the NASA/IPAC Extragalactic Database (NED) which is operated by the Jet Propulsion Laboratory, California Institute of Technology, under contract with the National Aeronautics and Space Administration.

\bibliographystyle{/Users/jorjales/Documents/TexDocs/Style_Files/mn2e.bst}
\bibliography{/Users/jorjales/Documents/TexDocs/Bibliography/1stpaper.bib}{}
\bibstyle

\label{lastpage}
\end{document}